\documentclass[conference]{IEEEtran}
\usepackage{CJKutf8}
\usepackage{comment}
\usepackage{amsmath}
\usepackage{tabularx}
\usepackage{booktabs}
\usepackage{multirow}
\usepackage{mathtools}
\usepackage{algorithm2e}
\usepackage{graphicx}
\usepackage{amssymb, mathtools}
\usepackage[hidelinks]{hyperref}
\usepackage{capt-of}
\usepackage[pass]{geometry}
\usepackage{xcolor}
\usepackage{balance}
\usepackage{CJKutf8}
\usepackage{float}
\usepackage{tabularx}

\input epsf
\usepackage{graphicx}

\begin{document}

%
\title{Origin-Destination Extraction from Large-Scale 
Route Search Records for Tourism Trend Analysis}



\author{
    \IEEEauthorblockN{Hangli Ge\IEEEauthorrefmark{1}, Dizhi Huang\IEEEauthorrefmark{2}, Xiaojie Yang\IEEEauthorrefmark{2}, Lifeng Lin\IEEEauthorrefmark{2}, Kazuma Hatano\IEEEauthorrefmark{1},
    Takeshi Kawasaki\IEEEauthorrefmark{3},
    Noboru Koshizuka\IEEEauthorrefmark{1}}
    \IEEEauthorblockA{
\IEEEauthorrefmark{1}Interfaculty Initiative in Information Studies; }
\IEEEauthorblockA{\IEEEauthorrefmark{2}Graduate School of Interdisciplinary Information Studies, The University of Tokyo\\
\IEEEauthorrefmark{3}Operation Division ITS Promotion Department, East Nippon Expressway Company Limited, Japan} \{hangli.ge,  dizhi.huang, xiaojie.yang, lifeng.lin, kazuma.hatano, noboru\}@koshizuka-lab.org
\IEEEauthorrefmark{3}t.kawasaki.ab@e-nexco.co.jp
}


\maketitle

\begin{abstract}
This paper presents a novel method for transforming large-scale historical expressway route search records into a three-dimensional (3D) Origin-Destination (OD) map, enabling data compression, efficient spatiotemporal sampling and statistical analysis. The study analyzed over 380 million expressway route search logs to investigate online search behavior related to tourist destinations. Several expressway interchanges (ICs) near popular attractions, such as those associated with spring flower viewing, autumn foliage and winter skiing, are examined and visualized. The results reveal strong correlations between search volume trends and the duration of peak tourism seasons. This approach leverages cyberspace behavioral data as a leading indicator of physical movement, providing a proactive tool for traffic management and tourism planning.
\end{abstract}


 
\section{Introduction}
Efficient traffic management has emerged as a critical concern for city planners and policymakers. It refers to the coordinated use of technology, infrastructure, and policy interventions to optimize the traffic flow. Its ultimate goals are to enhance safety, reduce traffic congestion, minimize environmental impact, and improve the quality of life for urban dwellers \cite{ristama2023utilization}. For instance, in recent years, certain regions and specific time periods have seen a high concentration of tourists, raising concerns about adverse impacts on the daily lives of local residents—particularly due to excessive traffic congestion—as well as a potential decline in the overall quality of the visitor experience.

Data mining offers a transformative approach to capture the traffic flow changes by extracting meaningful patterns, trends, and predictions from large and heterogeneous datasets. For instance, route search records on expressways accumulate user searches for routes, travel time, toll fees, and other related information, serving as a valuable resource for estimating future traffic demand. The monitoring of Origin-Destination (OD) data, route searches, and booking patterns can support predictive modeling and adaptive transportation planning for tourism regions. These insights can guide evidence-based decision-making for dynamic traffic control, and strategic planning in traffic management area \cite{hangli2022multi,ge2024k, ge2025traffic}. Especially when users specify departure or arrival times, these logs offer strong indicators for predicting future traffic trends. 
\begin{figure}[t]  
  \centering
   \includegraphics[width=0.98\linewidth]{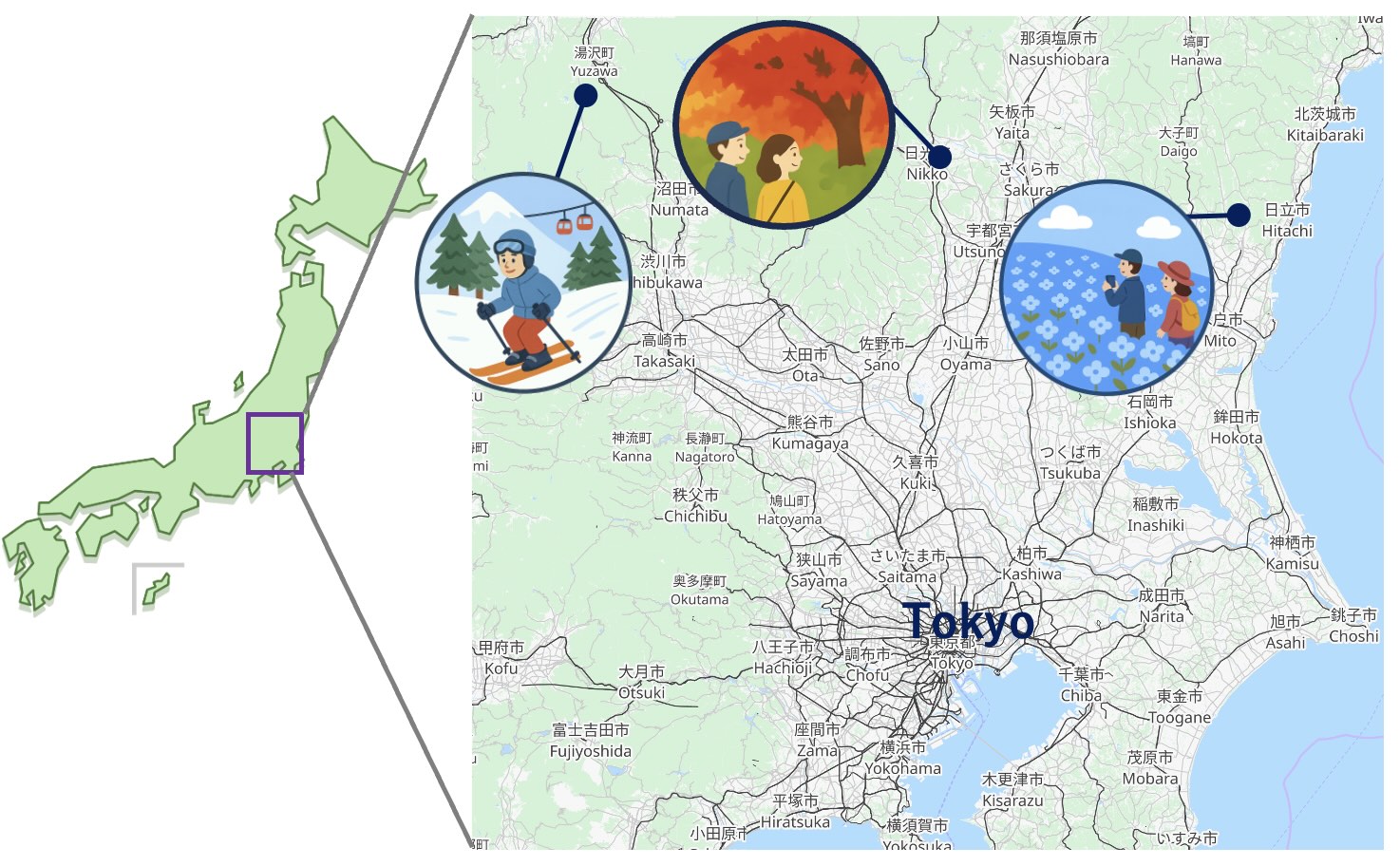} 
  \caption{The selected tourist areas for the case studies include nemophila flower viewing; autumn foliage viewing and winter ski tourism. The map used is based on OpenStreetMap.}
  \label{fig:fields}
\end{figure}


Building on this idea, this study proposes a novel method for transforming large-scale historical expressway route search records into a three-dimensional (3D) OD map. It supports data compression of large scale search logs, further enables efficient spatiotemporal sampling and statistical OD analysis of search volume. We utilize expressway route search logs from \textit{DoRaPuRa}, a toll and route search service operated by opreration division ITS promotion department, East Nippon Expressway Company Limited (NEXCO East, Japan) \footnote{\url{https://https://www.e-nexco.co.jp/en/}, accessed July-2025}.

Additionally, we conducted case studies focusing on expressway ICs located near popular tourist destinations that exhibit seasonal traffic trend patterns, including nemophila flower viewing in spring, foliage viewing in autumn, and ski tourism in winter (as illustrated in Figure~\ref{fig:fields}). By sampling data and visualizing route search volumes, our finding demonstrated that search data can effectively capture and reflect seasonal and periodical patterns in tourism demand. This study contributes to the fields of traffic forecasting and tourism analytics by introducing a novel approach that leverages cyberspace behavioral data as a proxy for physical traffic flow and human mobility. The proposed method offers valuable insights for traffic management, event planning, and personalized tourism promotion strategies.

\section{Datasets: Expressway Route Search Logs}
\begin{figure}[H]
  \centering
  \includegraphics[height=3.2in, width=0.8\linewidth]{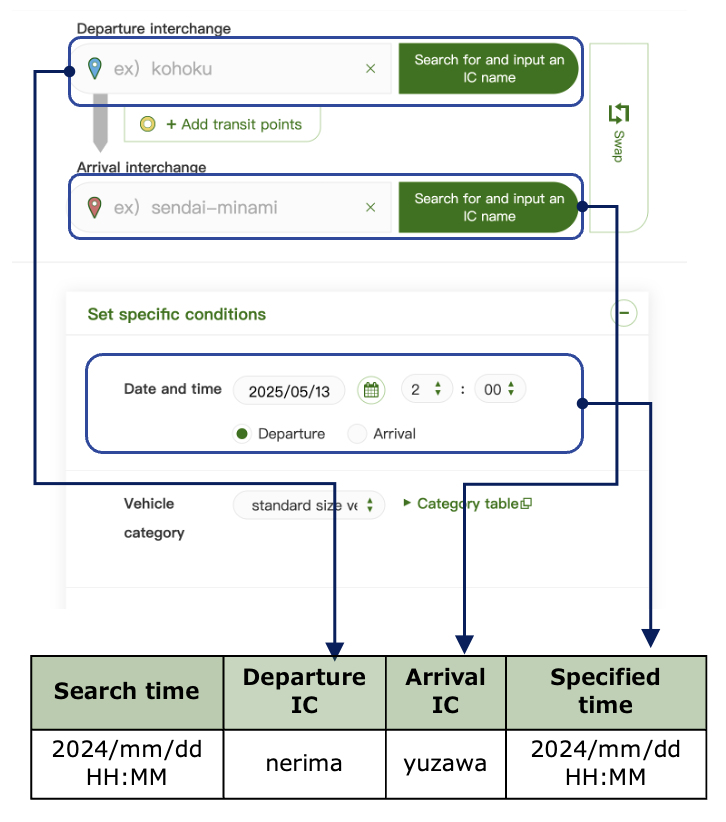}
  \caption{A screenshot showing the user interface of the DoRaPuRa service}
  \label{fig:dorapura}
\end{figure}

As the screenshot shown in Figure~\ref{fig:dorapura}, the \textit{DoRaPuRa} service \footnote{\url{https://en.driveplaza.com/}, accessed July-2025}.  allows users to input departure and destination interchanges (ICs) to inquiry information on routes, tolls, and travel times. If users do not specify a time for arrival or departure, the system automatically sets the time of website access as the default. Users may optionally specify departure or arrival times, vehicle types, and more. Accordingly, for this study, we extracted four key fields from each search: search timestamp, departure IC, arrival IC, specified time. We analyzed the search records spanning from April 2021 to August 2024, a total of 40 months and over 380 million entries. This corresponds to an average of approximately 309,000 searches per day. 


\begin{figure}[H] 
  \centering
  \includegraphics[height=1.8in, width=0.7\linewidth]{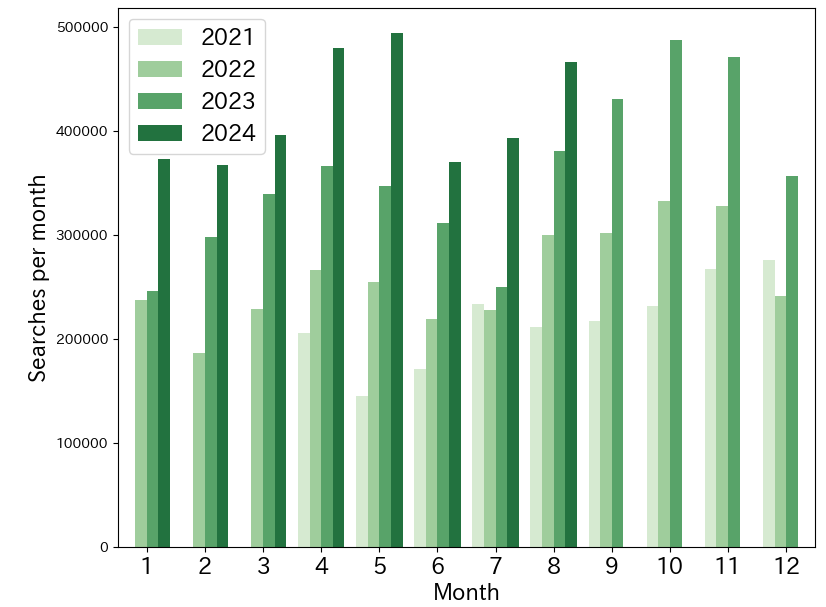}
  \caption{Monthly search volume trends from (April 2021 – August 2024)}
  \label{fig:monthly_trends}
\end{figure}

As shown in Figure ~\ref{fig:monthly_trends}, the monthly search volume trends indicate a year-on-year increase, with noticeable peaks in April, May, October, and November. These surges likely reflect seasonal events such as Golden Week and the autumn foliage. The highest search volume was recorded in May 2024, exceeding 49 million searches. In contrast, lower search activity was observed January, February, June, and July.


\section{Proposal of Three-Dimensional Origin–Destination (OD) Map}
\subsection{Preprocessing of Search Records}
\begin{figure}[h]
\centering 
\includegraphics[height=2.6in, width=0.45\textwidth]{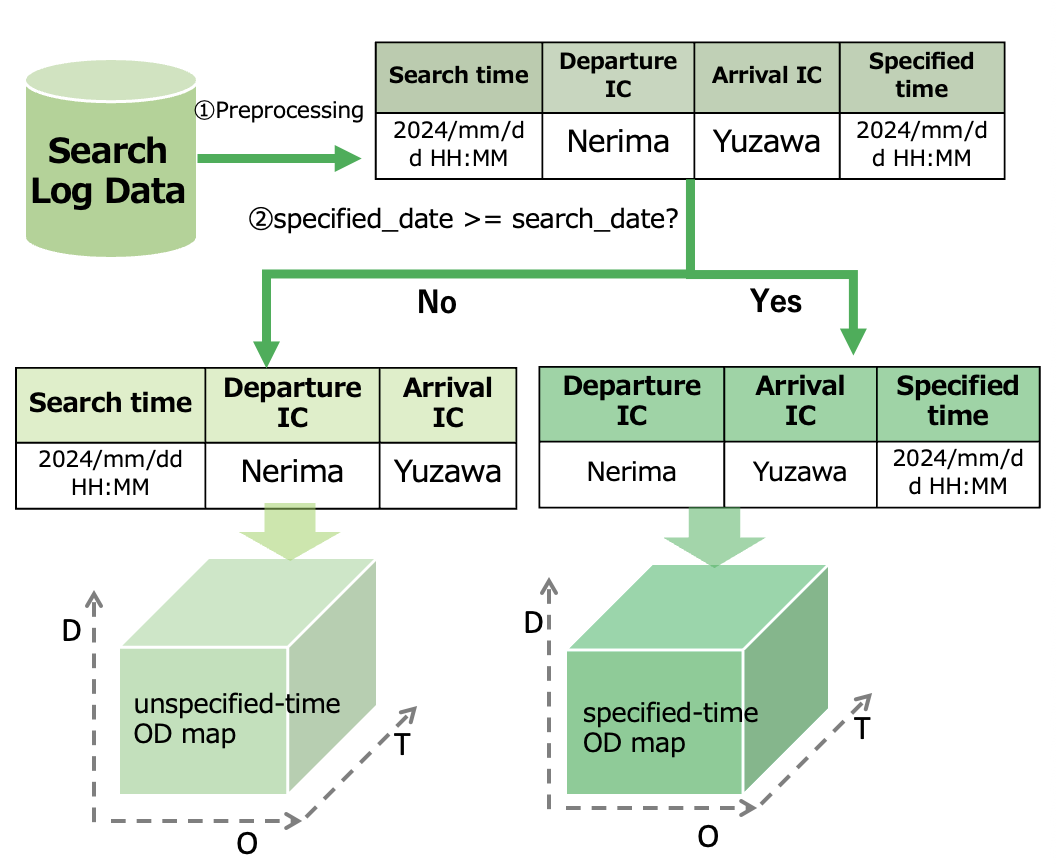} 
\caption{Search data processing flow} 
\label{fig5} 
\end{figure}
As indicated in Figure~\ref{fig5}, based on the value of specified time of \textit{DoRaPuRa} search logs, where the specified time precedes the search time are categorized as ``unspecified time" and others as ``specified time". Therefore, separated OD maps were built for each category. In this study, since the statistical processing was based on a daily basis, we classified the data into two categories: records in which the specified time is later than the search time were labeled as specified-time OD map, while all others were treated as unspecified-time OD map.

\subsection{Method for Constructing a 3D OD Map}
An OD map is a type of visualization used to represent origin-destination flows within a geographical context \cite{abrahamsson1998estimation,yang2016many,wood2010visualisation}. Traditionally, an OD map consists of a grid or matrix representation of flow data, with the geographic locations of origins and destinations displayed on the map. The size and color of cells or nodes can be used to represent the volume or intensity of flows between each origin-destination pair.

Building on the traditional OD map, we propose a novel construction method by adding time as a third dimension. In our proposal, search activity (denoted potential tourism volume) are represented in a 3D space of \( O \times D \times T \quad \text{(Origin × Destination × Days)} \), where the \( O \) and \( D \) axes denote search amount from origin to destination with quantitative variable, and the \( T\)-axis encodes a temporal variable as duration days. This enables efficient spatiotemporal analysis and trend extraction.

\begin{figure}[H]
\centering 
\includegraphics[height=1.2in, width=0.45\textwidth]{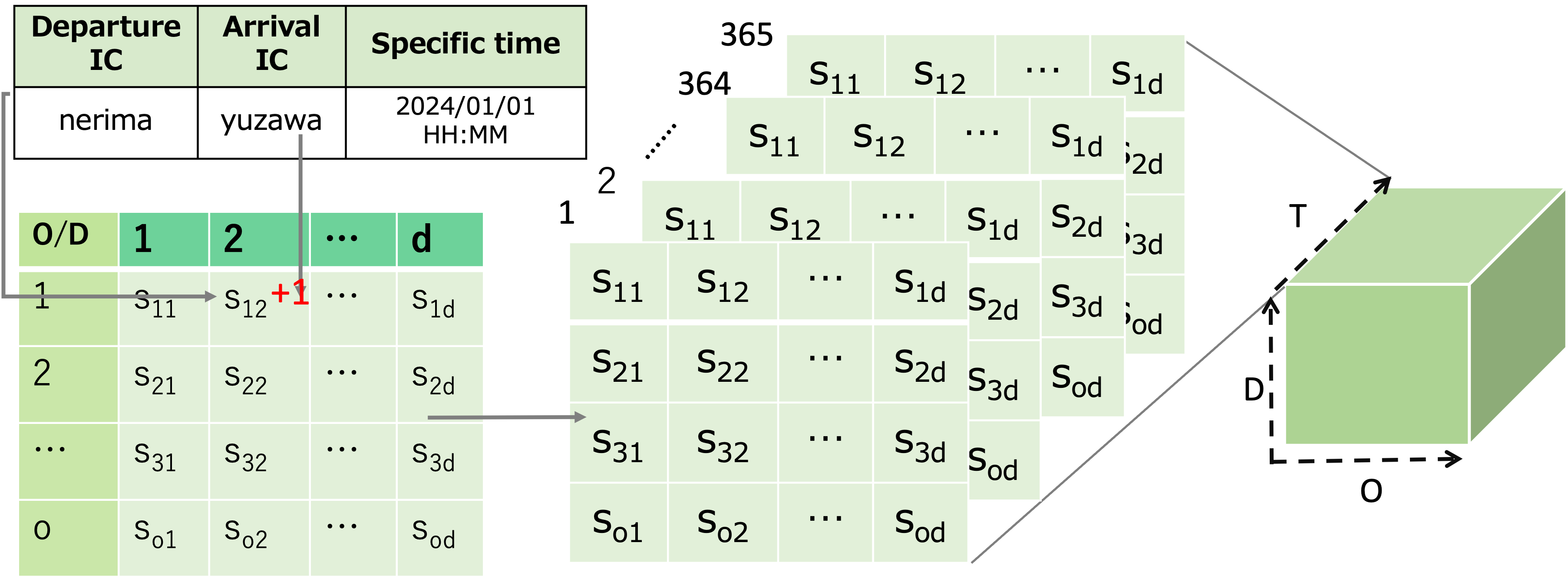} 
\caption{3D OD map construction} 
\label{img/fig3} 
\end{figure}

Figure~\ref{img/fig3} outlines the detailed process. First, ICs on the expresswaynetwork were assigned sequential numbers based on longitude and latitude, resulting in 2,728 ICs. It forms a two dimensional OD matrix of size 2728 x 2728 initialized with zeros. Daily statistical processing was performed, with one two dimensional OD map generated per day. These are stacked over the year, creating a 3D OD map with dimensions  \( O \times D \times T \quad \text{(Origin × Destination × Days)} \). Each log entry is processed to calculate the number of days elapsed from January 1st, based on the specified time (or search time if unspecified). Using the departure IC (\(i\)) and destination IC (\(j\)), the corresponding element \(S_{i,j}\) in the OD matrix for the calculated search volume is incremented by one. Repeating this process for all logs completes the 3D OD map.

\subsection{Advantages of constructing 3D OD Map}
\textbf{Compressing Large-Scale Data}: By transforming unstructured and voluminous search logs into a structured \( O \times D \times T \) format, we achieved effective data compression. It enables more efficient downstream processing and visualization. Python libraries such as NumPy and Pandas were used, and the resulting data was stored as NumPy 3D arrays. The raw data (approx. 4.6 GB) was divided into two datasets (with and without time specification), compressed to 290MB and 141MB respectively.
\begin{figure}[H]
\centering 
\includegraphics[height=1in, width=0.6\linewidth]{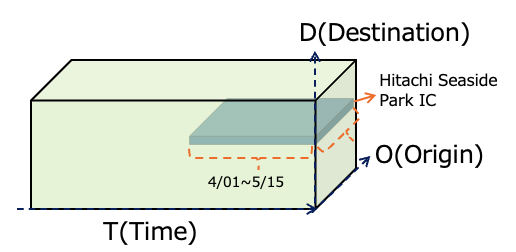} 
\caption{Example of extracting search statistics of specific IC} 
\label{fig2} 
\end{figure}
\textbf{Efficient Analysis of IC-Level Statistics}: Using NumPy library, we can quickly slice and analyze data. For instance, np.sum(3d\_od, axis=0) aggregates across the departure axis, yielding total searches to each destination IC. Filtering by destination IC or by time range (e.g., April 1 to May 15 corresponds to day index from 90 to 136) is done via slicing such as 3d\_od[:, :, 90:136] (as an example illustrated in Figure~\ref{fig2}). Similarly, sorting origin ICs by search volume for a given destination IC can be performed with np.sort() or np.argsort(), enabling extraction of top-ranked OD pairs for congested tourist areas.

\section{Case Studies}
\begin{figure*}
  \centering
  \includegraphics[height=3.5in, width=0.96\linewidth]{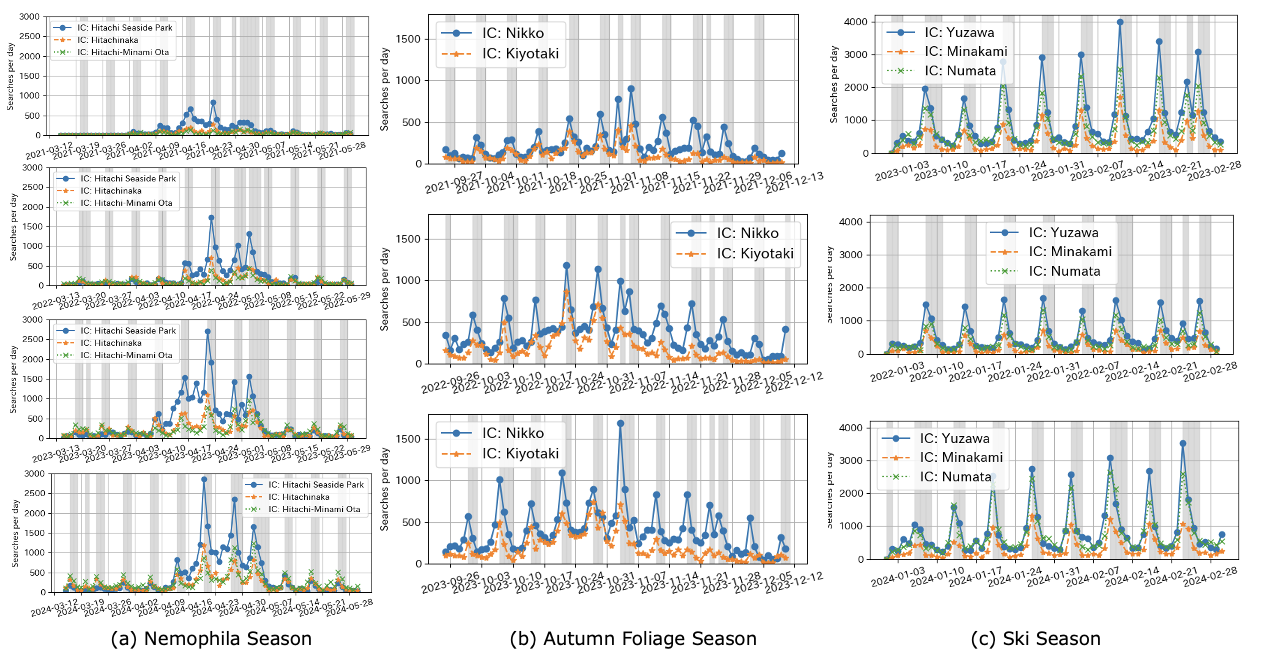}
  \caption{Number of searches during the tourist season: weekend was highlighted in grey-colored background}
  \label{fig:3case}
\end{figure*}
In the context of transportation and tourism analysis, we evaluated event-driven traffic fluctuations. We conducted case studies on three seasonal tourism scenarios: nemophila viewing in spring around Hitachi area, autumn foliage viewing around Nikko area, and winter ski tourism around Yuzawa area (Figure~\ref{fig:fields}). Expressway ICs near the relevant sites were selected and analyzed using the 3D OD map. As shown in Figure~\ref{fig:3case}, search volumes increased year by year, indicating a recovery in tourism demand post-COVID. Peaks frequently occurred on weekends and public holidays, suggesting that tourists tend to plan trips on such days. Consistent trends were observed across ICs during the same season, revealing common seasonal patterns.

\subsection{Nemophila Season}
Nemophila, commonly known as baby blue eyes, is a spring-blooming flower. It typically occurs from mid-April to early May, peaking around Japan’s Golden Week holidays. Among the most famous viewing sites is Hitachi Seaside Park in Ibaraki Prefecture, where over 5 million Nemophila flowers bloom \cite{hitachi_nemophila}. We visualized time-specified search volumes from March 15 to May 30  for each year from 2021 to 2024, focusing on the following ICs: Hitachi Seaside Park, Hitachinaka, and Hitachi-Minami Ota. As shown in Figure~\ref{fig:3case} (a), search volumes increased annually during the nemophila season, particularly on weekends preceding Golden Week. This pattern suggests that tourists may be trying to avoid Golden Week holiday-related congestion.

\subsection{Autumn Foliage Season}
Nikko, located in Tochigi Prefecture, is one of Japan’s most renowned destinations for viewing autumn foliage. The area is characterized by its mountainous terrain, historical temples and shrines and natural beauty, making it a prime location for autumn foliage  tourism. The autumn foliage season typically begins in early to mid-October at higher elevations. From the perspective of tourism and traffic analytics, Nikko’s autumn foliage period serves as a key case study for event-driven mobility, allowing for the analysis of origin–destination (OD) patterns, seasonal peak load management. We analyzed search volumes from September 25 to December 10 for each year from 2021 to 2023, focusing on the ICs near popular foliage destinations: Nikko and Kiyotaki. As shown in Figure~\ref{fig:3case}(b), search volumes peaked in mid-October to early November. While Kiyotaki remained stable across years, Nikko showed a noticeable increase in 2023.

\subsection{Ski Season}
Yuzawa, located in Niigata Prefecture, is one of Japan’s most accessible and popular winter resort areas, particularly known for its long ski season. From a transportation and tourism data perspective, Yuzawa provides a valuable case study in seasonal travel behavior. During the ski season, the region experiences a notable spike in transportation demand, evident in increased train occupancy, expressway congestion, and accommodation bookings.  For ski destinations (January 1 to March 31 for each year from 2021 to 2023), we analyzed search volumes for ICs near Yuzawa, Minakami, and Numata. As shown in Figure ~\ref{fig:3case}(c), Yuzawa had the highest search volume, followed by Minakami and Numata. This aligns with high demand for ski resorts around these areas.




\section{Conclusion and Future Work}
This study proposed a novel method for constructing a three-dimensional OD map from large-scale expressway route search logs by incorporating time as a third dimension. Case studies on seasonal tourism revealed that search data effectively reflect real-world travel demand. Cyberspace data, such as search logs and reservations can support congestion mitigation, targeted outreach, and long-term traffic forecasting. Our findings also apply to related domains like long-term traffic prediction \cite{kosugi2022traffic,matsunaga2023improving,ge2024frtp}. Accurately capturing this surge through route search data or mobility logs can offer valuable insights into managing temporary congestion, traffic distribution \cite{ge2025traffic}, and regional promotion strategies.
We aim to further enhance the integration of cyberspace big data into real-world decision support systems.


\section{Acknowledgement}
This study was conducted under a joint research project between the University of Tokyo and the East Nippon Expressway Co., Ltd. (NEXCO East). This study was also supported by JSPS KAKENHI Grant Number JP25K21205. The data used in this study, such as historical online search logs, and road structure information, were provided by NEXCO East. We gratefully acknowledge the kind support provided by the NEXCO East.


\balance
\bibliography{IEEEabrv,aaai22}

\bibliographystyle{IEEEtran}

\end{document}